\documentclass[aps,prd,twocolumn,floatfix,nofootinbib]{revtex4}

\usepackage[T1]{fontenc}
\usepackage[utf8]{inputenc}
\usepackage{lmodern}
\usepackage{amsmath,amssymb,mathtools}
\usepackage{graphicx}
\usepackage{booktabs}
\usepackage{array}
\usepackage{multirow}
\usepackage{caption}
\usepackage{hyperref}
\usepackage{siunitx}
\usepackage{microtype}
\usepackage{xcolor}
\hypersetup{colorlinks=true,linkcolor=blue,citecolor=blue,urlcolor=blue}
\graphicspath{{./}}
\begin{document}

\title{On trace invariance and energy-dependent entanglement in muon decay}

\author{Saulo Carneiro}
\affiliation{Observat\'orio Nacional, 20921-400 Rio de Janeiro, RJ, Brazil}

\author{Fernando César Sobrinho}
\affiliation{Instituto de F\'isica, Universidade de S\~ao Paulo, 05508-090 S\~ao Paulo, SP, Brazil}

\begin{abstract}
We revisit a proposed correction to the muon magnetic anomaly measurement and compare it with energy-binned scans reported by the Fermilab $g-2$ collaboration. The central point is that, for a relativistic three-body decay inferred from an ensemble of detected positrons, the relevant neutrino partial trace is defined on the asymptotic state at the detector, not on an idealised state at the decay vertex. Because the fitted positron sample mixes events with different decay times, flight times, energies and acceptances, tracing over the unobserved neutrino helicities leaves a time-dependent mixed state in the observed sector. Using angular-momentum balance, we obtain an energy-dependent correction in the positron energy interval $E_p \in (1.5,2.9)$~GeV. The corrected experimental interval for the muon anomaly moves towards the BaBar- and $\tau$-based determinations, while remaining compatible at the $1\sigma$ level with both data-driven and lattice-based Standard Model evaluations. We then fit digitised public $R(E_p)$ scans for Fermilab's runs $1$ to $6$, with two one-parameter hypotheses: a constant shift and the entanglement-motivated energy-dependent shift. Within digitisation accuracy, 
the differences in $\chi^2$ remain modest and the current binned data do not provide decisive discrimination between them.
\end{abstract}


\maketitle

\section{Introduction}

The muon anomalous magnetic moment remains a high-precision consistency test of the Standard Model and, simultaneously, a probe of the connection between theory and experiment \cite{Aoyama2020,Aliberti2025}. On the experimental side, Brookhaven E821 and Fermilab E989 determine the anomalous precession frequency from a time-modulated positron signal \cite{Bennett2006,Albahri2021,Aguillard2023}. On the theory side, the dominant uncertainty is the hadronic vacuum polarisation, whose evaluation currently splits into a data-driven dispersive programme and lattice-QCD determinations \cite{Aoyama2020,Borsanyi2021,Aliberti2025}.

In two previous papers we proposed that part of the remaining discrepancy may originate in the experimental inference itself \cite{BJP2023,RE2026}. The proposal does not alter QED, electroweak or hadronic loop calculations, but it questions whether the standard map from the measured positron modulation to the inferred muon anomaly is complete once the angular momentum carried by the unobserved neutrino pair is taken into account. In the first article the effect was expressed in terms of proper Larmor frequencies, while in the second it was reformulated in the lab frame, based on angular-momentum balance.

In the present paper we reinforce the conceptual point that the relevant neutrino trace operation is attached to the asymptotic detected state rather than to an idealised state at the decay vertex. We confront the predicted energy dependence of the effect with public energy-binned scans extracted from four theses produced by the Muon $g-2$ Collaboration \cite{Girotti2023,Foster2023,LaBounty2024,Israel2026}. Within the accuracy allowed by the available figures, we show that an entanglement-motivated correction is favoured, while the current binned data do not distinguish it decisively from a constant shift.

\section{The Fermilab experiments}

The experimental observable is the time distribution of detected decay positrons. A polarised beam of positive muons is stored in the magnetic ring, the spin precesses relative to the momentum, and the parity-violating decay $\mu^+\to e^+\nu_e\bar{\nu}_\mu$ transfers the muon-spin information to the positron angular distribution \cite{Tiomno1949,Michel1950,Bennett2006,Albahri2021}. 

In practice, however, the fitted quantity is not a single-event observable. It is reconstructed from a weighted ensemble that depends on calorimeter response, pileup subtraction, gain corrections, lost-muon modelling, asymmetry weighting, fit windows, and the ratio method \cite{Albahri2021,Aguillard2024,Foster2023,Girotti2023}. These features show that the experiment infers $\omega_a$ from a structured statistical ensemble rather than from an ideal one-particle measurement.

The interval \SIrange{1.5}{2.9}{GeV} is the natural window for the present comparison. Below about \SI{1.5}{GeV} the analysing power becomes weak and the uncertainties grow rapidly. Above that threshold the bins retain substantial sensitivity and any additional energy dependence can be confronted with the published scans \cite{BJP2023,Girotti2023,Foster2023,LaBounty2024,Israel2026}.

\section{The tension between data-driven and lattice results}

The tension in muon $g-2$ is no longer a simple theory-versus-experiment one. It is also a competition between two ways of computing the hadronic vacuum polarisation. The data-driven programme uses dispersive integrals based on $e^+e^-\to$ hadrons cross sections, whereas the lattice programme aims at an \emph{ab initio} QCD determination \cite{Aoyama2020,Borsanyi2021,Aliberti2025}.

Lattice QCD is conceptually clean, yet it depends on a chain of numerical approximations: continuum extrapolation, finite-volume corrections, chiral behaviour at the physical point, treatment of disconnected diagrams, isospin-breaking effects, and the matching between Euclidean-window observables and the full HVP contribution \cite{Borsanyi2021,Aliberti2025}. None of these ingredients is fatal in itself, but the issue is cumulative. A lattice number is only as convincing as the stability of the entire extrapolation and matching procedure under changes of action, lattice spacing, volume and analysis window.

The data-driven route is closer to experiment but inherits the systematics of the underlying cross sections. In the two-pion channel, the recent CMD-3 result disagrees with older $e^+e^-$ datasets \cite{CMD32024,KLOE2018,Aliberti2025}, forcing renewed scrutiny of radiative corrections, event selection and normalisation because of the size of its impact on $a_\mu$. By contrast, the BaBar ISR data remain attractive because of their internal consistency and broad control of radiative effects \cite{BaBar2009,Davier2020}, while $\tau$-based evaluations bring an independent weak-current perspective, although they require carefully modelled isospin-breaking corrections before they can be compared with $e^+e^-$ data \cite{Davier2020}. The present situation therefore requires a careful assessment of the input data, because the various channels do not currently carry the same empirical weight.

From the standpoint of the present analysis, the proposed correction does not resolve the data-driven versus lattice discrepancy by itself. Its effect is instead to shift the experimental point downward by an amount that depends on the positron-energy window, thereby moving the corrected result towards the region favoured by BaBar- and $\tau$-based evaluations, while remaining statistically compatible with both the 2020 data-driven white-paper value and the 2025 lattice-based update \cite{Aoyama2020,Aliberti2025,RE2026}.

\section{The partial trace}

The standard criticism of the entanglement proposal is that, once the neutrino degrees of freedom are traced out, any subsequent neutrino evolution must drop out of the reduced positron signal. In ordinary finite-dimensional quantum mechanics this would be the default expectation. In the present relativistic setting, however, that conclusion requires additional assumptions \cite{Peres2004,RE2026}.

The muon decay is a relativistic three-body process in an external magnetic field,
and the experiment never measures an ideal state at the decay vertex. What it measures is an ensemble of asymptotic events, positrons selected with different flight times, decay positions, acceptances, energies and weights in the final fit \cite{Aguillard2024,Foster2023,Girotti2023}. The trace that enters the real observable is therefore the trace over the unobserved sector in the detector-defined out state. It is not a formal trace over a fictitious subsystem frozen at the vertex \cite{RE2026}.

In other words, the positron sample used to infer $g$ is not a pure state, it is a time-dependent mixed ensemble. Each detected positron belongs to a decay with its own proper-time history, and the modulation extracted from the ensemble is a weighted average over subensembles whose composition changes through time \cite{Aguillard2024,Foster2023}.

The missing neutrino sector carries away not only energy and momentum but also angular momentum information \cite{Daniel}. Once those degrees of freedom are traced over, the surviving positron ensemble is left with a mixed angular structure. The invariance of the trace does not authorise one to ``de-evolve'' that mixed state back to the decay vertex as if one were dealing with a nonrelativistic subsystem decomposition. In relativistic quantum theory only the asymptotic {\it in} and {\it out} states are operationally clean \cite{Peres2004,Landau 4}.

Within this interpretation, the additional modulation is not a retrocausal signal from the neutrino sector to the positron sector. It is a correction to the inference map that connects the detected positron modulation to the underlying muon anomaly.

\section{Angular momentum, neutrino precession and the anomaly correction}

Our quantitative estimate starts from the angular momentum carried by the neutrino pair. Consider that they are emitted in the $x$-direction, with initial longitudinal polarisation, and that the magnetic field lies along the $z$-direction. Following \cite{RE2026}, the $y$-component of the neutrino angular momentum in the laboratory frame is \cite{Landau 2}
\begin{equation}
L_y = \gamma_\nu \left( s_y - v_\nu L'^{03} \right),
\end{equation}
with
\begin{equation}
L'^{03}= t' p'_z - E' z'
\end{equation}
in the neutrino rest frame. Under precession in the external magnetic field, its infinitesimal change is
\begin{equation}
\delta L_y = -\gamma_\nu \mu_\nu B\,dt,
\end{equation}
where $\mu_\nu$ is the neutrino magnetic moment. 

Since the decay produces both a neutrino and an antineutrino moving forward in the lab frame, the uncertainty transferred to the parent-muon angular momentum due to neutrino tracing is doubled. Equating the neutrino-side and muon-side variations yields
\begin{equation}
\delta \mu_\mu\,\gamma_\mu = 2\mu_\nu\,\gamma_\nu,
\end{equation}
and therefore, for the muon magnetic anomaly,
\begin{equation}
\delta a_\mu = \left( \frac{2m_\mu}{e}\right) \delta \mu_\mu \approx 7.2\,f\times 10^{-9}.
\end{equation}
Here we have adopted the magnetic moment of a Dirac massive neutrino \cite{Giunti,PDG},
\begin{equation}
    \mu_{\nu} = \frac{3eG_Fm_{\nu}}{8\sqrt{2}\pi^2},
\end{equation}
and $f = 2m_\nu \gamma_\nu / (m_\mu \gamma_\mu)$ is the fraction of the muon energy carried by the neutrino pair. 

In the simplest kinematic reading, $f$ increases when the detected positron carries less energy. In practice, however, the experimental sensitivity is modulated by the positron asymmetry, which grows with $E_p$. The competition between these two tendencies produces the characteristic hump-shaped correction discussed in our first paper \cite{BJP2023}.
The relevant average over energy bins is weighted with the energy-dependent asymmetry \cite{BJP2023,Bennett2006,Aguillard2024},
\begin{equation} \label{media}
\bar{\delta a}_\mu = \frac{\int dE_p\,A(E_p)\,\delta a_\mu(E_p)}{\int dE_p\,A(E_p)},
\end{equation}
where the asymmetry $A(E_p)$ is the effective analysing power entering the fitted modulation\footnote{The experimental sensitivity can also be modelled by weighting the average with $NA^2$. The event distribution is, however, dominated by low-energy positrons, which carry lower sensitivity. In the experimental $T$-method, this is controlled by imposing run-dependent lower energy thresholds. An estimate using the inverse squared uncertainties of the energy-binned scans as a proxy for $NA^2$ increases the averaged anomaly correction by about $15\%$.}.

\begin{figure}[t]
    \includegraphics[width=\columnwidth]
    {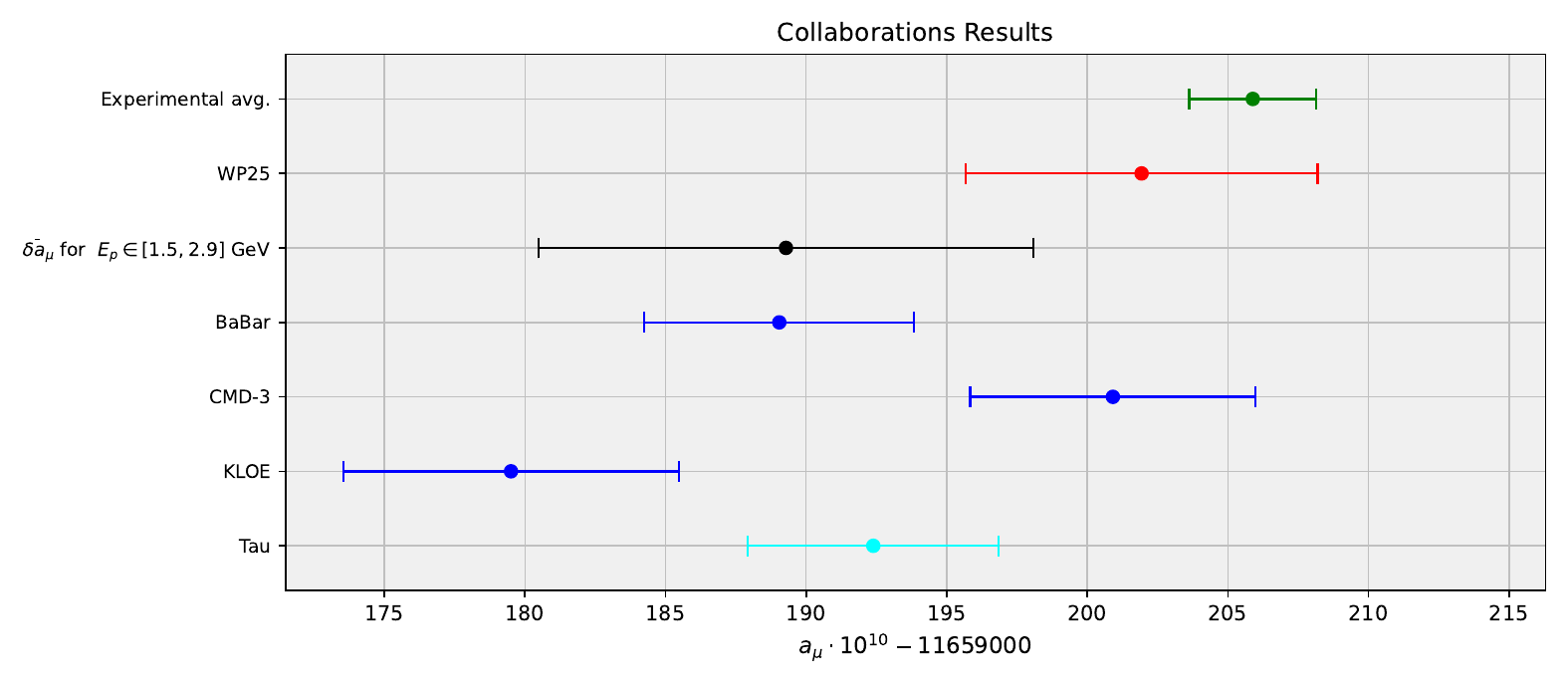}
    \caption{The full black bar is the $1\sigma$ interval for the corrected magnetic anomaly, the green bar the uncorrected experimental average, blue and cyan the data-driven evaluations, and red the lattice-based result. The black-bar interval combines the standard uncertainty associated with the average in (\ref{media}) with the experimental $1\sigma$ interval.}
    \label{fig:bar-summary}
\end{figure}

Using the energy window that dominates the public high-precision scans, the average correction is of order \SI{1.4}{ppm}. Figure~\ref{fig:bar-summary} gives a comparison chart, adapted from our previous paper \cite{RE2026}. After averaging over \SIrange{1.5}{2.9}{GeV}, the corrected experimental interval shifts from the uncorrected one towards the sector occupied by the BaBar- and $\tau$-based evaluations. At the same time, the interval continues to overlap the uncertainty bands associated with both the data-driven 2020 white-paper results and the lattice-based 2025 update \cite{Aoyama2020,Aliberti2025}. The correction therefore does not remove the hadronic ambiguity. Rather, it lowers the experimental point into the range where the two theoretical estimates overlap.

\section{The energy-binned scans}

The main new contribution of the present paper is the comparison with energy-binned scans available in four theses of the Muon $g-2$ Collaboration \cite{Girotti2023,Foster2023,LaBounty2024,Israel2026}. Initially, we digitised the $R(E_p)$ points from the published figures for the Run-1A panel from Girotti's thesis \cite{Girotti2023}, and for the Run-2, Run-3a and Run-3b panels from Foster's thesis \cite{Foster2023}. 

Our fits to the Foster data with a constant anomaly shift, in the entire observed interval of energies, give best-fits and quality indicators in precise agreement with those reported in the thesis, which supports confidence in the digitisation precision. Nevertheless, because
the corresponding tables and covariance matrices are not publicly available, the numerical estimates below should be interpreted with caution rather than as substitutes for a full collaboration reanalysis. Even so, the results show that the energy-dependence predicted by the entanglement hypothesis cannot be falsified within the current experimental precision.

\begin{figure}[t]
    \centering
    \includegraphics[width=.85\columnwidth]{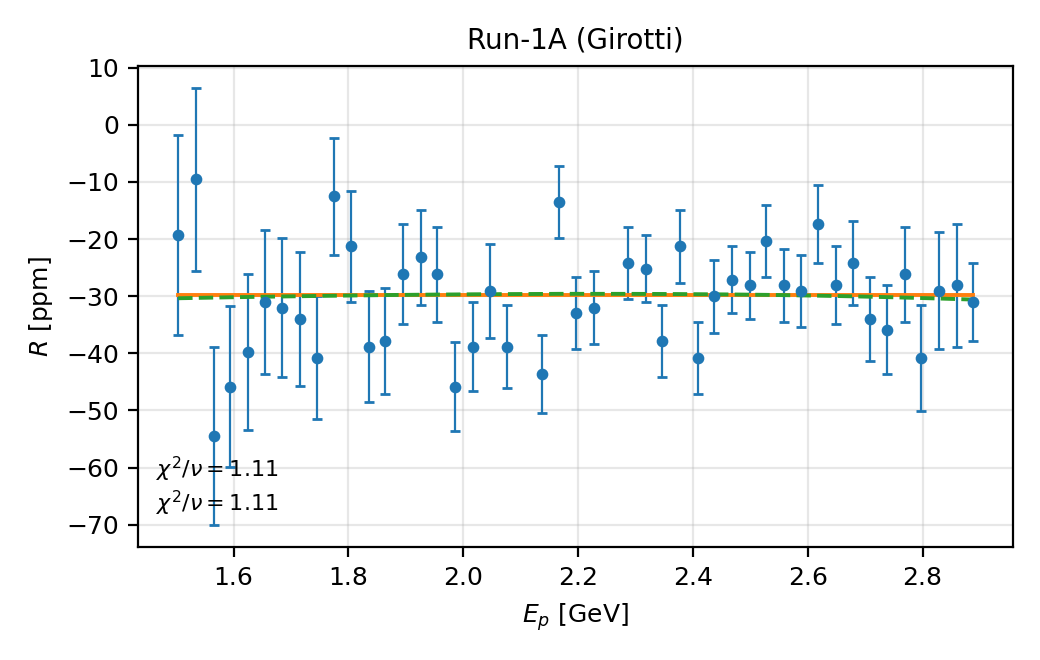}
    \caption{Digitised Run-1A scan obtained from Fig.~7.4(c) of Girotti's thesis \cite{Girotti2023}. The curves were recomputed from the selected energy bins using the same two one-parameter models used for the Foster panels: a constant shift and an entanglement-motivated energy-dependent shift.}
    \label{fig:girotti-fit}
\end{figure}

\begin{figure}[t]
    \centering
    \includegraphics[width=\columnwidth]{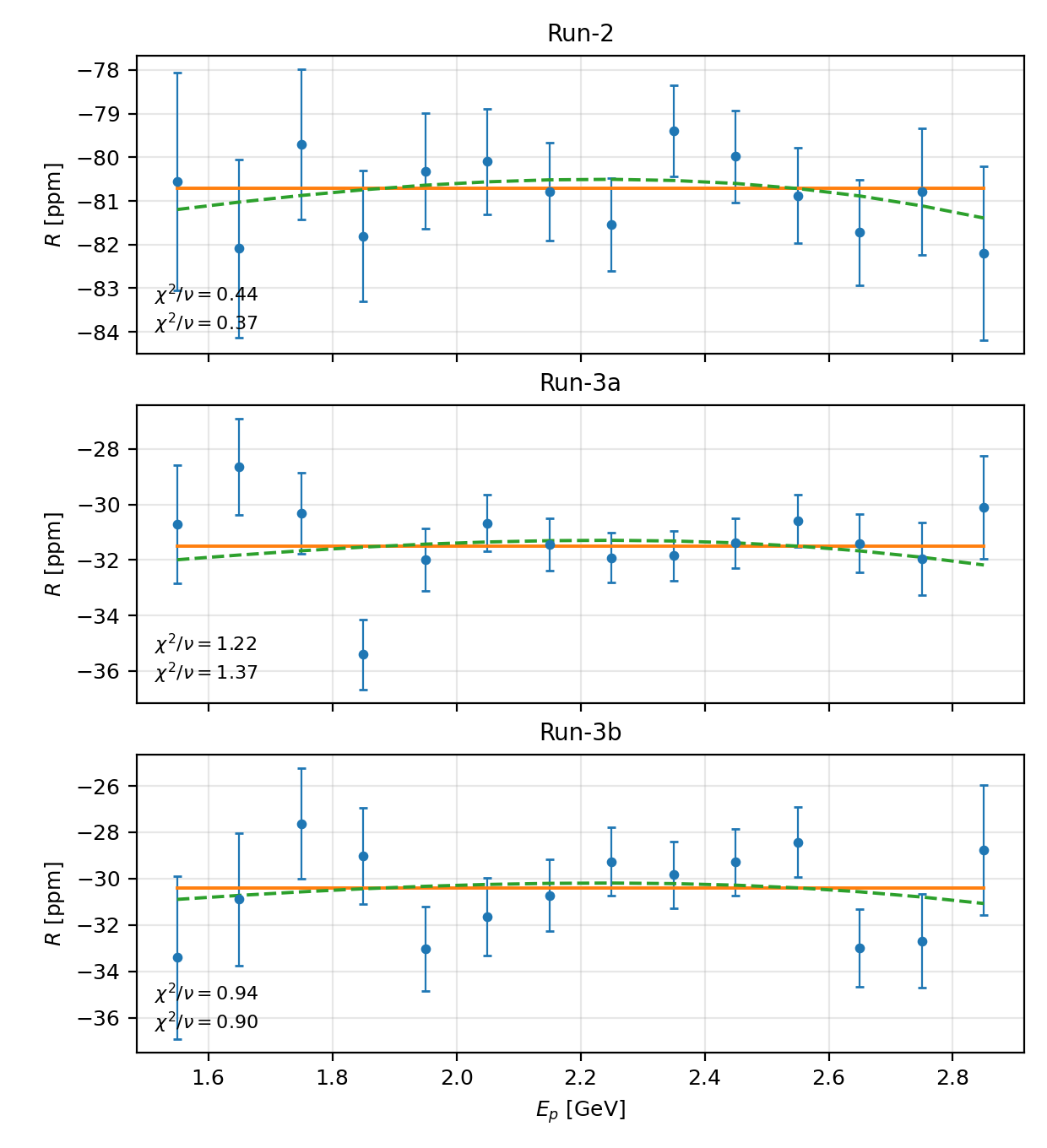}
    \caption{Digitised Run-2, Run-3a and Run-3b scans from Foster \cite{Foster2023}. Blue points are the selected digitised data. The orange solid curves are the recomputed constant fits, and the green dashed curves are the energy-dependent fits.}
    \label{fig:foster-fits}
\end{figure}

We compare two one-parameter models. In the constant model the blinded variable $R$ is represented by a single horizontal offset. In the entanglement model the offset is replaced by the energy profile implied by our correction \cite{BJP2023}, with no extra shape parameter added in the fit.
Figure~\ref{fig:girotti-fit} shows the Run-1A fit obtained after digitising Fig.~7.4(c) of Girotti's thesis \cite{Girotti2023}. Figure~\ref{fig:foster-fits} shows the corresponding fits for the Run-2, Run-3a and Run-3b panels in Foster's thesis \cite{Foster2023}.

\begin{table}[t]
    \caption{Fit-quality comparison for the digitised energy-binned scans. Both constant ($c$) and energy-dependent ($e$) models use the same number of fitted parameters.}
    \label{tab:fits}
    \small
    \setlength{\tabcolsep}{3.5pt}
    \begin{ruledtabular}
    \begin{tabular}{lrr}
        Dataset &  $(\chi^2/\nu)_{\rm c}$ & $(\chi^2/\nu)_{\rm e}$ \\
        \midrule
        Run-1A (Girotti) & 1.11 & 1.11 \\
        Run-2 (Foster)   & 0.44 & 0.37 \\
        Run-3a (Foster)  & 1.22 & 1.37 \\
        Run-3b (Foster)  & 0.94 & 0.90 \\
    \end{tabular}
    \end{ruledtabular}
\end{table}

Table~\ref{tab:fits} shows the corresponding quality indicators after refitting the selected bins. The Run-1A values are essentially identical for the two hypotheses. Run-2 and Run-3b give marginally smaller values for the energy-dependent profile, whereas Run-3a favours the constant-shift hypothesis. These differences remain small compared with the limitations imposed by digitisation and by the absence of public covariance matrices. Overall, the binned data do not provide a decisive model selection.

More recent energy-binned scans are available in the theses of LaBounty and Israel \cite{LaBounty2024,Israel2026}. These studies partly analyse the same runs $4-6$ events with different estimators and therefore cannot be treated as independent measurements. For the explicit comparison below we retain Israel's T-method scans for the four running-condition groups Run-noRF, Run-xRF, Run-5xyRF and Run-6xyRF. The corresponding RT-method scans use essentially the same events and are not included. The LaBounty scans provide useful consistency checks, but adding them to Table~\ref{tab:runs456-fixed-fits} would repeat data already represented by the selected runs 4--6 groups.

The four groups follow the experimental running conditions rather than the calendar run alone. In particular, Run-xRF and Run-5xyRF correspond to distinct periods of Run~5: the former uses horizontal RF, whereas the latter uses both horizontal and vertical RF. They are therefore separate acquisition samples, not duplicated analyses of the same events.

For the validation of our digitisation of the Israel panels, we used the subseries directly constrained by the numerical information reported in the thesis. Although additional points are visible in the panels, the constant-fit labels in Fig.~5.23 quote numbers of degrees of freedom compatible with a smaller effective set of points, while Table~5.7 reports the weighted averages \(\langle R(E)\rangle\) obtained from $1$ GeV to $3$ GeV in $100$ MeV steps. Therefore, for each running-condition group, we selected for validation the subseries whose full-range constant fit reproduces both the reported \(\langle R(E)\rangle\) and the degrees of freedom printed in the figure. The other visible points, however, are not necessarily spurious and were all included in the fits below, consistently with the $50$ MeV binning stated for the energy-binned fits.

\begin{figure*}[t]
    \centering
    \includegraphics[width=0.45\textwidth]{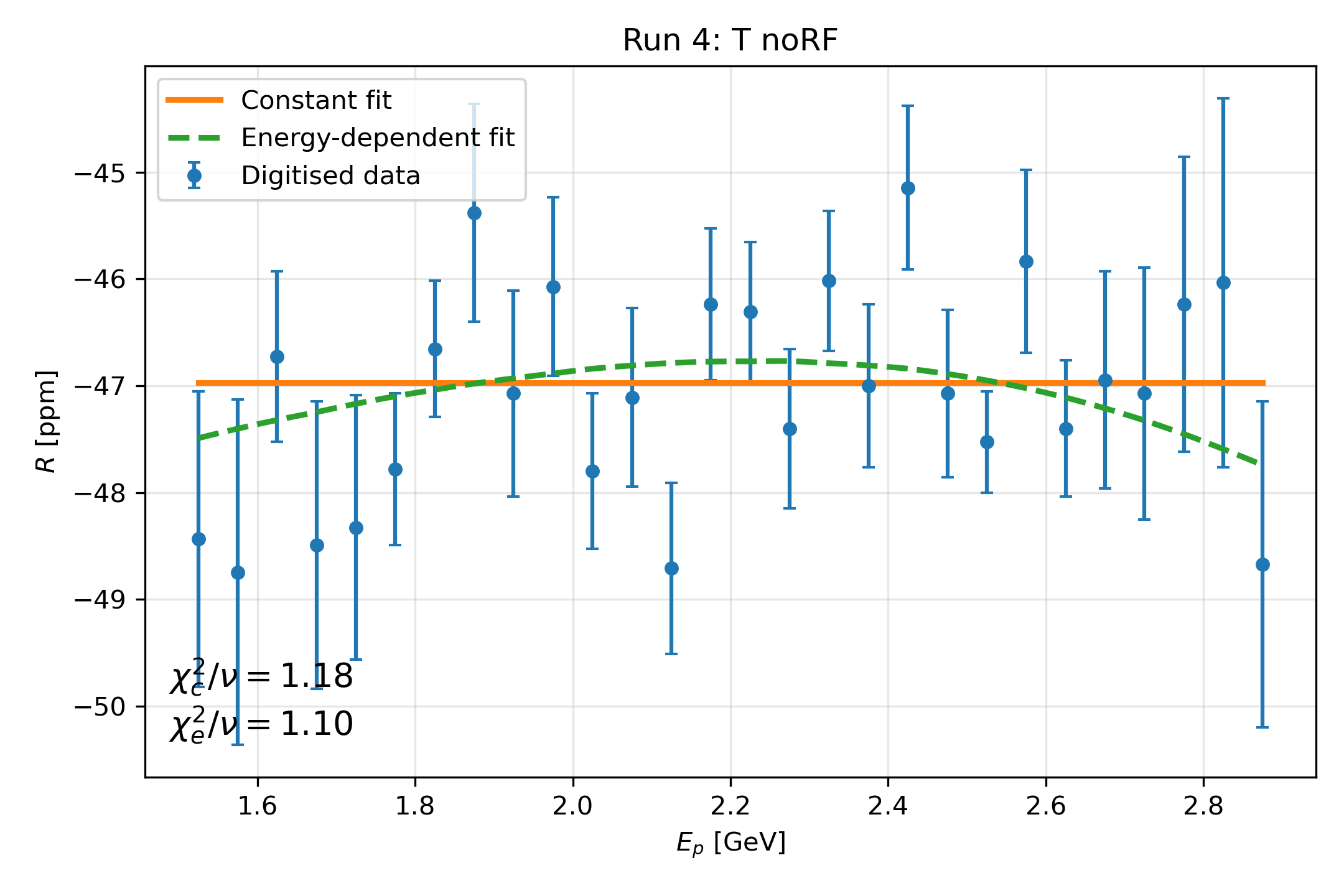}\hfill
    \includegraphics[width=0.45\textwidth]{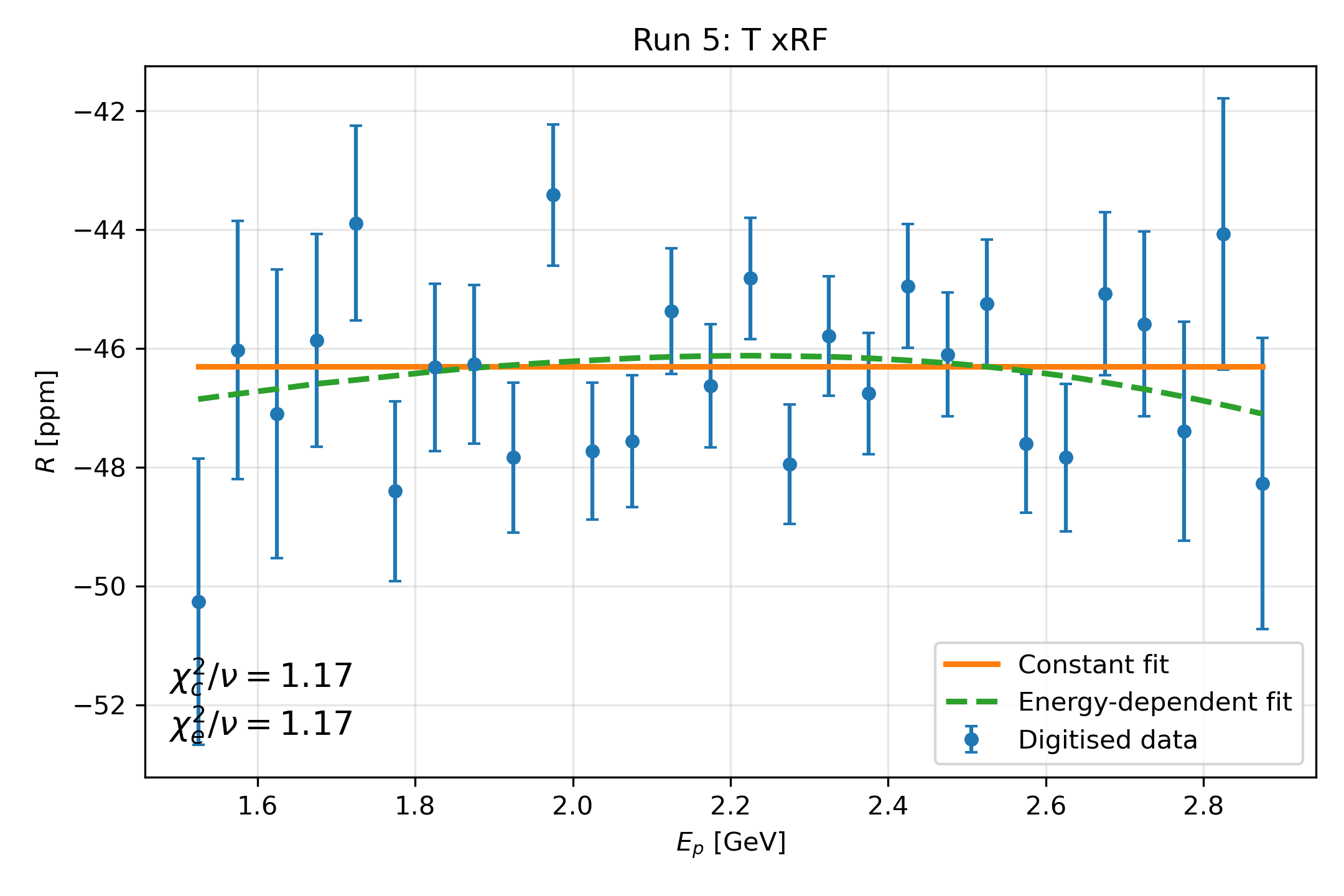}

    \vspace{5pt}
    \includegraphics[width=0.45\textwidth]{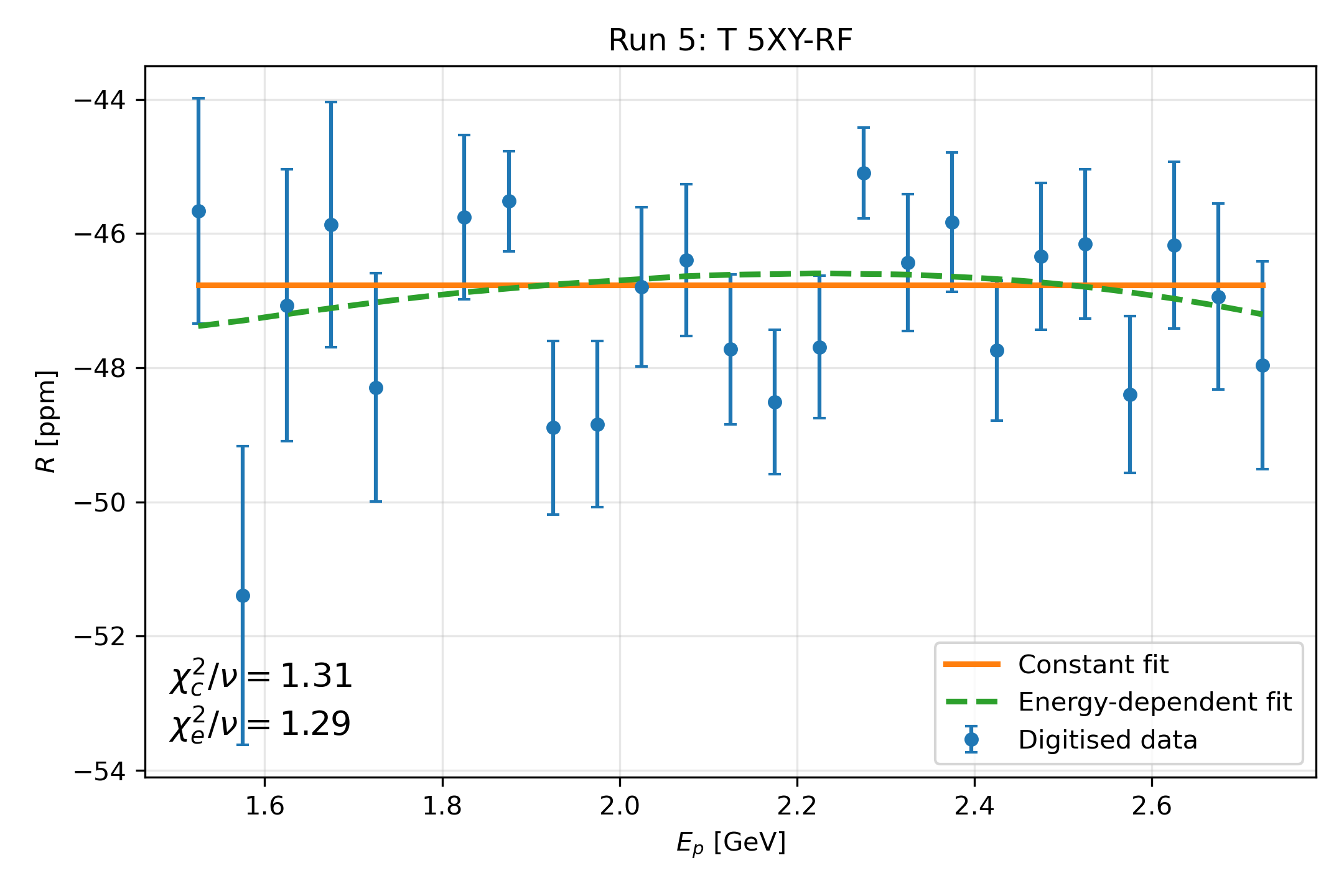}\hfill
    \includegraphics[width=0.45\textwidth]{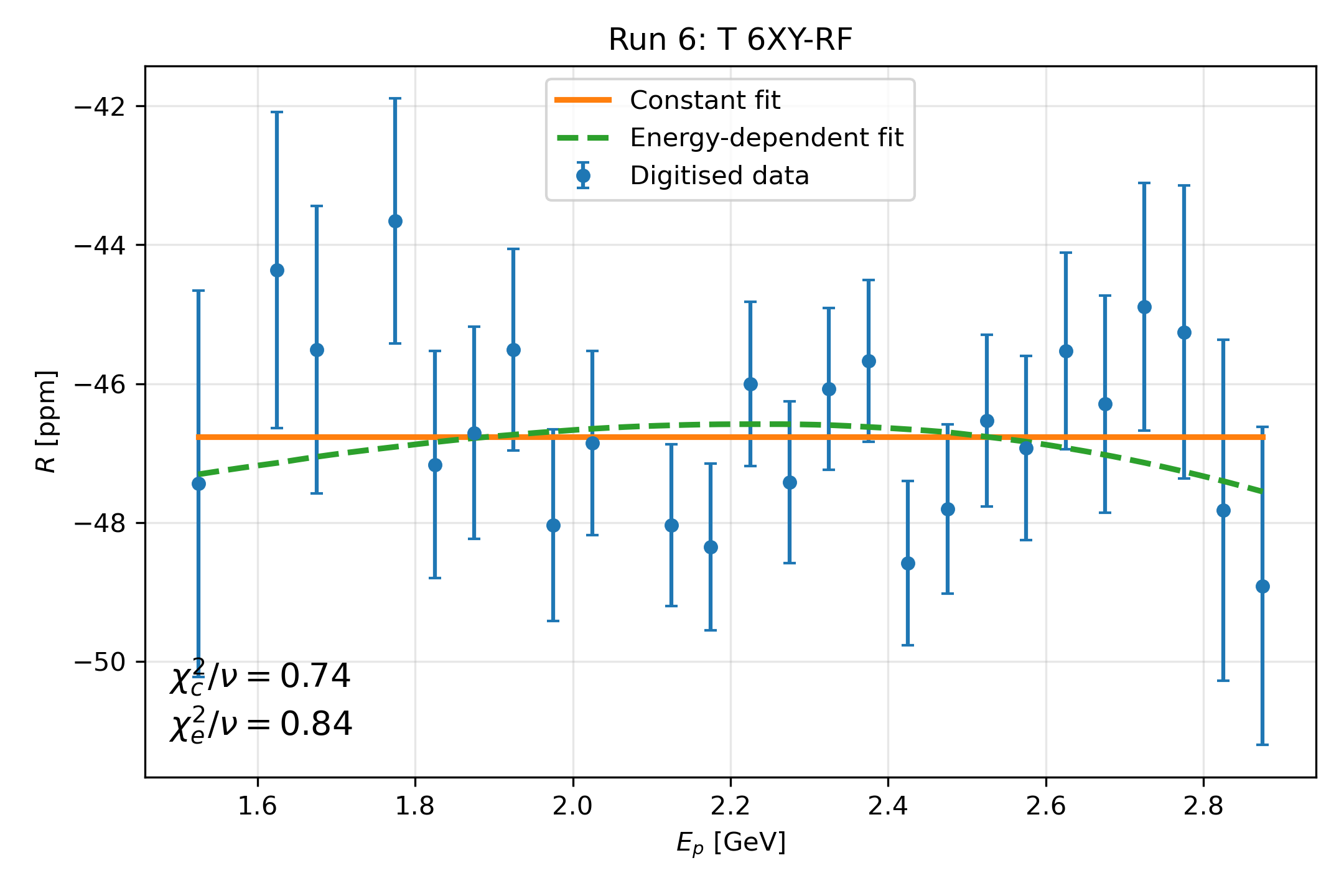}
    \caption{Digitised T-method energy-binned scans for the four runs 4--6 groups from Israel \cite{Israel2026}. Blue points are the digitised points used in the present fits. The orange solid curves are the constant fits, and the green dashed curves are the energy-dependent fits. The RT-method scans are omitted because they are constructed from essentially the same events.}
    \label{fig:runs456-fixed-fits}
\end{figure*}

\begin{table}[t]
    \caption{Fit-quality comparison for the selected runs $4-6$ scans. As in Table~\ref{tab:fits}, both the constant ($c$) and energy-dependent ($e$) models contain the same number of fitted parameters.}
    \label{tab:runs456-fixed-fits}
    \small
    \setlength{\tabcolsep}{4.5pt}
    \begin{ruledtabular}
    \begin{tabular}{lcc}
        Dataset & $(\chi^2/\nu)_{\rm c}$ & $(\chi^2/\nu)_{\rm e}$ \\
        \midrule
        Run-noRF  & 1.18 & 1.10 \\
        Run-xRF   & 1.17 & 1.17 \\
        Run-5xyRF & 1.31 & 1.29 \\
        Run-6xyRF & 0.74 & 0.84 \\
    \end{tabular}
    \end{ruledtabular}
\end{table}
The selected scans from runs $4$--$6$ give acceptable fit qualities for both models, with no uniform preference for either hypothesis. For the Israel scans, the values in Table~\ref{tab:runs456-fixed-fits} were recomputed using the $50$ MeV binning. For Run-xRF the quality indicators are both acceptable. Run-noRF and 
Run-5xyRF marginally favour the energy-dependent shift in the selected window, and Run-6xyRF remains below unity in both cases. This pattern does not, by itself, select between the two physical hypotheses. The resulting fits are shown in Table~\ref{tab:runs456-fixed-fits} and in Figure~\ref{fig:runs456-fixed-fits}.

A global and more general test of the energy-dependence hypothesis can also be done, by fitting the whole dataset with the ansatz
\begin{equation}
\omega_a = \omega_{a0} \left( 1 + c \frac{\delta a_\mu}{a_\mu} \right),
\end{equation}
which interpolates the two above hypotheses, the constant case for $c = 0$ and the energy-dependent one for $c = 1$. In the above expression, $\delta a_\mu$ is weighted with the positron asymmetry associated with each scan \cite{BJP2023}.

The coefficient $c$ should not be interpreted as an arbitrary rescaling of a sharply predicted correction. The neutrino angular-momentum contribution defines a maximum uncertainty scale, while the neutrino Larmor frequency represents only a characteristic frequency of the additional modulation reconstructed from a heterogeneous positron ensemble. Accordingly, $c=1$ corresponds to maximal coherent transfer, whereas $0<c<1$ describes partial projection of that uncertainty onto the experimentally fitted modulation.

The common coefficient $c$ was then allowed to vary in the global fit, with an independent normalisation for each scan. For $c$ we have adopted the uniform prior $c \in [-2,2]$, and for the positron energy range we have adopted the interval $1.5$ GeV $< E_p < 2.9$ GeV as before, for which the error bars and dispersions are reliable. No selection based on the individual fit quality was applied. 

The global fit gives
\begin{equation}
    c=0.527 \pm 0.450\quad(68\%\ \text{c.l.}),
\end{equation}
with
\begin{equation}
    \chi^2/\nu=1.055.
\end{equation}
The energy-independent case $c=0$ and the maximal-profile case $c=1$ 
lie, respectively, $1.17\sigma$ and $1.05\sigma$ from the best fit.
Both hypotheses provide acceptable absolute fits, and the improvement at the best fit is modest. The result is therefore marginally compatible with a statistical interpretation, but it motivates a dedicated analysis using the original energy-binned data and their covariance matrices.

\section{Conclusions}

The argument developed in our previous papers may be summarised in the following way \cite{BJP2023,RE2026}. The muon $g-2$ experiment does not measure a pure positron state at the decay vertex. It reconstructs a detector-defined mixed ensemble of positrons with different decay times, flight times, energies, acceptances and asymmetry weights \cite{Aguillard2024,Foster2023,Girotti2023}. In that setting, tracing over the unobserved neutrino sector can leave a residual incompleteness in the standard inference from the detected positron modulation to the muon anomaly. When the angular momentum carried by the neutrino pair is estimated, the resulting correction is naturally energy dependent and numerically of the same order as the long-standing anomaly in the muon $g-2$ factor \cite{RE2026}.

At the phenomenological level, the correction moves the experimental interval towards the BaBar- and $\tau$-based determinations while maintaining compatibility, within $1\sigma$ confidence level, with both data-driven and lattice-based Standard Model evaluations \cite{Aoyama2020,Aliberti2025}. At methodological level, the analysis highlights that the trace operation relevant for the experiment is attached to the asymptotic observed state and not to an idealised state at the decay vertex. In this relativistic setting, no trace invariance theorem authorises a de-evolution of the detector-level out state to the vertex \cite{RE2026,Peres2004,Landau 4}.

The comparison with energy-binned scans is the main new result of the present paper. After digitising the runs $1-6$ panels extracted from \cite{Girotti2023,Foster2023,LaBounty2024,Israel2026}, we find that the currently public data do not discriminate robustly between an energy-dependent correction and a constant shift. This was already the situation for the Brookhaven points considered in our earlier article \cite{BJP2023}, and it remains true for the Fermilab thesis material examined here. 

The best-fit value found for $c$ in the global fitting of the different scans, lying between the two limiting hypotheses, should not be simply read as an arbitrary phenomenological interpolation. It rather may suggest that the energy-dependent modulation associated with the neutrino angular-momentum contribution is only partially projected onto the experimentally reconstructed positron ensemble. This is natural, since the fitted $\omega_a$ is obtained from a mixture of events with different energies, decay times, acceptances and analysis weights.

The natural next step would be the release of
$R(E_p)$ points together with the corresponding covariance matrices. 
Only with that information can the present hypothesis be tested in a fully controlled way.

\section*{Acknowledgements}

\noindent FCS thanks FAPESP (São Paulo, Brazil) for his PhD grant number 2024/19103-9. SC is supported by CNPq with grant number 308518/2023-3.

\end{document}